\documentclass[12pt]{elsart}
\usepackage{epsfig}
\usepackage{pstricks}

\def\b{\beta}

\def\e{\epsilon}

\def\j{\psi}
\def\k{\kappa}

\def\m{\mu}
\def\n{\nu}

\def\p{\pi}

\def\t{\tau}

\def\D{\Delta}
\def\F{\Phi}
\def\G{\Gamma}

\def\L{\Lambda}

\def\Q{\Theta}

\def\ve{\varepsilon}

\def\cl{{\mathcal L}}

\def\co{{\mathcal O}}

\def\cv{{\mathcal V}}

\def\bo{{\raise.15ex\hbox{\large$\Box$}}}               % D'Alembertian
\def\pr{\prod}                                          % product
\def\face{{\raise.2ex\hbox{$\displaystyle \bigodot$}\mskip-2.2mu \llap {$\ddot
        \smile$}}}                                      % happy face
\def\dg{\dagger}                                     % hermitian conjugate
\def\wt#1{\widetilde{#1}}                    % big tilde
\def\Bar#1{\overline{#1}}                       % big bar
\def\VEV#1{\left\langle #1\right\rangle}        % < >
\def\leftrightarrowfill{$\mathsurround=0pt \mathord\leftarrow \mkern-6mu
        \cleaders\hbox{$\mkern-2mu \mathord- \mkern-2mu$}\hfill
        \mkern-6mu \mathord\rightarrow$}       % <--> double differential
\def\dvec#1{\vbox{\ialign{##\crcr
        \leftrightarrowfill\crcr\noalign{\kern-1pt\nointerlineskip}
        $\hfil\displaystyle{#1}\hfil$\crcr}}}           % <--> accent
\def\beq{\begin{equation}}
\def\eeq{\end{equation}}

\def\beqx{\begin{displaymath}}
\def\eeqx{\end{displaymath}}

\def\beqa{\begin{eqnarray}}
\def\eeqa{\end{eqnarray}}
\def\NO{\nonumber}

\def\pl#1#2#3{Phys.~Lett.~{\bf B {#1}} (19{#2}) #3}
\def\np#1#2#3{Nucl.~Phys.~{\bf B {#1}} (19{#2}) #3}
\def\prl#1#2#3{Phys.~Rev.~Lett.~{\bf #1} (19{#2}) #3}
\def\pr#1#2#3{Phys.~Rev.~{\bf D {#1}} (19{#2}) #3}

\def\prep#1#2#3{Phys.~Rep.~{\bf {#1}C} (19{#2}) #3}

\def\nc#1#2#3{Nuovo Cim.~{\bf {#1}} (19{#2}) #3}

\catcode`@=11
\def\@citex[#1]#2{\if@filesw\immediate\write\@auxout{\string\citation{#2}}\fi
  \def\@citea{}\@cite{\@for\@citeb:=#2\do
    {\@citea\def\@citea{,\penalty\@m}\@ifundefined
      {b@\@citeb}{{\bf ?}\@warning
       {Citation `\@citeb' on page \thepage \space undefined}}%
\hbox{\csname b@\@citeb\endcsname}}}{#1}}

\def\citer{\@ifnextchar [{\@tempswatrue\@citexr}{\@tempswafalse\@citexr[]}}
\def\@citexr[#1]#2{\scriptsize 
  \if@filesw\immediate\write\@auxout{\string\citation{#2}}\fi
  \def\@citea{}\@cite{\@for\@citeb:=#2\do
    {\@citea\def\@citea{-\penalty\@m}\@ifundefined
       {b@\@citeb}{{\bf ?}\@warning
       {Citation `\@citeb' on page \thepage \space undefined}}%
\hbox{\csname b@\@citeb\endcsname}}}{#1}\normalsize}
\catcode`@=12

\begin{document}

\begin{frontmatter}
\title{
{\normalsize
\textnormal{DESY 99-044} \hspace{\fill} \mbox{ }\\
\textnormal{UPR-0845-T} \hspace{\fill} \mbox{ }\\
\textnormal{April 1999} \hspace{\fill} \mbox{ }\\[5ex]
}
MATTER ANTIMATTER ASYMMETRY AND NEUTRINO PROPERTIES\thanksref{okun}}
\thanks[okun]{Contribution to the Festschrift for L.~B.~Okun, 
to appear in a special issue of Physics Reports, 
eds.~V.~L.~Telegdi and K.~Winter}
\author{Wilfried~Buchm\"uller}
\address{Deutsches Elektronen-Synchrotron DESY, Hamburg, Germany}

\author{Michael~Pl\"umacher}
\address{Department of Physics and Astronomy, University of Pennsylvania,\\
Philadelphia PA 19104, USA}

\begin{abstract}
\noindent
The cosmological baryon asymmetry can be explained as remnant of heavy
Majorana neutrino decays in the early universe. We study this mechanism
for two models of neutrino masses with a large $\n_\m-\n_\t$ mixing angle
which are based on the symmetries $SU(5)\times U(1)_F$ and 
$SU(3)_c\times SU(3)_L\times SU(3)_R\times U(1)_F$, respectively.
In both cases $B-L$ is broken at the unification scale $\L_{GUT}$.
The models make different predictions for the baryogenesis
temperature and the gravitino abundance.
\end{abstract}

\end{frontmatter}

\newpage

\section{Baryogenesis and lepton number violation}

The cosmological matter antimatter asymmetry, the ratio of the baryon density 
to the entropy density of the universe,
\beq
  Y_B = {n_B-n_{\Bar{B}}\over s} = (0.6 - 1)\cdot 10^{-10}\;,
  \label{blub}
\eeq
can in principle be understood in theories where baryon number, C and
CP are not conserved \cite{sac}. The
presently observed value of the baryon asymmetry is then explained as a 
consequence of the spectrum and interactions of elementary particles, 
together with the cosmological evolution.

A crucial ingredient of baryogenesis is the connection between baryon number
($B$) and lepton number ($L$) in the high-temperature, symmetric phase of
the standard model. Due to the chiral nature of the weak interactions $B$ and
$L$ are not conserved\cite{thoo}. At zero 
temperature this has no observable effect due to the smallness of the weak 
coupling. However, as the temperature approaches the critical temperature 
$T_{EW}$ of the electroweak phase transition, $B$ and $L$ violating processes 
come into thermal equilibrium\cite{krs}. These `sphaleron processes' violate 
baryon and lepton number by three units,
\beq 
    \D B = \D L = 3\;. \label{sphal1}
\eeq
It is generally believed that $B$ and $L$ changing processes are in thermal 
equilibrium for temperatures in the range
\beq 
T_{EW} \sim 100\ \mbox{GeV} < T < T_{SPH} \sim 10^{12}\ \mbox{GeV}\;.
\eeq
  
The non-conservation of baryon and lepton number has a profound effect on the 
generation of the cosmological baryon asymmetry.  
Eq.~\ref{sphal1} suggests that any
$B+L$ asymmetry generated before the electroweak phase transition,
i.e., at temperatures $T>T_{EW}$, will be washed out. However, since
only left-handed fields couple to sphalerons, a non-zero value of
$B+L$ can persist in the high-temperature, symmetric phase if there
exists a non-vanishing $B-L$ asymmetry. An analysis of the chemical potentials
of all particle species in the high-temperature phase yields the following
relation between the baryon asymmetry $Y_B$ and the corresponding
$L$ and $B-L$ asymmetries $Y_L$ and $Y_{B-L}$, respectively\cite{chem},
\beq\label{basic}
Y_B\ =\ a\ Y_{B-L}\ =\ {a\over a-1}\ Y_L\;,
\eeq
where $a$ is a number ${\mathcal O}(1)$. In the standard model with three 
generations and two Higgs doublets one has $a=8/23$. 
  
We conclude that $B-L$ violation is needed if the baryon asymmetry is
generated before the electroweak transition, i.e. at temperatures 
$T > T_{EW} \sim 100$~GeV. In the standard model, as well as its 
supersymmetric version and its unified 
extensions based on the gauge group SU(5), $B-L$ is a conserved quantity. 
Hence, no baryon asymmetry can be generated dynamically in these models.

The remnant of lepton number violation at low energies is 
an effective $\Delta L=2$ interaction between lepton and Higgs fields,
  \beq\label{dl2}
  \cl_{\Delta L=2} = {1\over 2} f_{ij}\
  l_{Li}^T H_2\ C\ l_{Lj} H_2 +\mbox{ h.c.}\;.\label{intl2}
  \eeq
Such an interaction arises in particular from the exchange of heavy Majorana
neutrinos. In the Higgs phase of the standard model, where the Higgs field 
acquires a vacuum expectation value $\VEV{H_2}=v_2$, it gives
rise to Majorana masses of the light neutrinos $\n_e$, $\n_\m$ and $\n_\t$.   

At finite temperature the $\Delta L=2$ processes described by (\ref{dl2}) take
place with the rate\cite{fy1}
  \beq
    \Gamma_{\Delta L=2} (T) = {1\over \pi^3}\,{T^3\over v_2^4}\, 
    \sum_{i=e,\m,\t} m_{\n_i}^2\; .
  \eeq
In thermal equilibrium this yields an additional relation between the
chemical potentials which implies
\beq
Y_B\ =\ Y_{B-L}\ =\ Y_L\ =\ 0 \; .
\eeq
To avoid this conclusion, the $\Delta L=2$ interaction (\ref{intl2}) must not 
reach thermal equilibrium. For baryogenesis at a temperature
$T_B < T_{SPH} \sim 10^{12}$ GeV, one has to require 
$\G_{\D L=2} < H|_{T_B}$, where $H$ is the Hubble parameter. This yields a 
stringent upper bound on Majorana neutrino masses,
\beq\label{nbound}
\sum_{i=e,\m,\t} m_{\n_i}^2 < \left(0.2\;\mbox{eV}\,
\left({T_{SPH}\over T_B}\right)^{1/2}\,\right)^2\;.
\eeq
For $T_B \sim T_{SPH}$, this bound would be comparable to the upper bound 
on the electron neutrino mass obtained from neutrinoless double beta decay. 
However, eq.~(\ref{nbound}) also applies to the $\t$-neutrino mass. 
Note, that the bound can be evaded if appropriate asymmetries are present 
for particles which reach thermal equilibrium only at temperatures below 
$T_B$ \cite{cli93}.

The connection between lepton number and the baryon asymmetry is lost
if baryogenesis takes place at or below the Fermi scale\cite{dol}. However, 
detailed studies of the thermodynamics of the electroweak transition have
shown that, at least in the standard model, the deviation from thermal
equilibrium is not sufficient for baryogenesis\cite{jansen}. In the minimal 
supersymmetric extension of the standard model (MSSM) such a scenario 
appears still possible for a limited range of parameters\cite{dol}.

\section{Decays of heavy Majorana neutrinos}

Baryogenesis above the Fermi scale requires $B-L$ violation, and therefore 
$L$ violation. Lepton number violation is most simply realized by adding 
right-handed Majorana neutrinos to the standard model.  Heavy right-handed 
Majorana neutrinos, whose existence is predicted by all extensions of the
standard model containing $B-L$ as a local symmetry, can also explain the 
smallness of the light neutrino masses via the see-saw mechanism\cite{seesaw}.

The most general Lagrangian for couplings and masses of charged
leptons and neutrinos reads 
\beq\label{yuk}
\cl_Y = -h_{e ij}\Bar{e_R}_i l_{Lj} H_1 
        -h_{\n ij}\Bar{\n_R}_i l_{Lj} H_2
        -{1\over2}h_{r ij} \Bar{\n^c_R}_i \n_{Rj} R
            +\mbox{ h.c.}\;.
\eeq
The vacuum expectation values of the Higgs field $\VEV{H_1}=v_1$ and
$\VEV{H_2}=v_2=\tan{\b}\ v_1$ generate Dirac masses $m_l$ and $m_D$ for 
charged leptons and neutrinos, $m_e=h_e v_1$ and $m_D=h_{\n}v_2$, 
respectively, which are assumed to be much smaller than the Majorana masses 
$M = h_r\VEV{R}$.
This yields light and heavy neutrino mass eigenstates
  \beq
     \n\simeq K^{\dg}\n_L+\n_L^c K\quad,\qquad
     N\simeq\n_R+\n_R^c\, ,
  \eeq
with masses
  \beq
     m_{\n}\simeq- K^{\dg}m_D{1\over M}m_D^T K^*\,
     \quad,\quad  m_N\simeq M\, .
     \label{seesaw}
  \eeq
  Here $K$ is a unitary matrix which relates weak and mass eigenstates. 
  
  The right-handed neutrinos, whose exchange may erase any lepton
  asymmetry, can also generate a lepton asymmetry by means of
  out-of-equilibrium decays. This lepton asymmetry is then partially 
  transformed into a baryon asymmetry by sphaleron processes\cite{fy}.  
  The decay width of the heavy neutrino $N_i$ reads at tree level,
  \beq
    \G_{Di}=\G\left(N_i\to H_2+l\right)+\G\left(N_i\to H_2^c+l^c\right)
           ={1\over8\p}(h_\n h_\n^\dg)_{ii} M_i\;.
    \label{decay}
  \eeq
From the decay width one obtains an upper bound on the light neutrino masses
via the out-of-equilibrium condition\cite{fisch}. Requiring 
$\Gamma_{D1}< H|_{T=M_1}$ yields the constraint
\beq\label{ooeb}
\wt{m}_1\ =\ (h_\n h_\n^\dg)_{11} {v_2^2\over M_1}\ < \ 10^{-3}\, \mbox{eV}\;.
\eeq
More direct bounds on the light neutrino masses depend on the structure
of the Dirac neutrino mass matrix.

Interference between the tree-level amplitude and the one-loop 
self-energy and vertex corrections yields  $CP$ asymmetries in the heavy
Majorana neutrino decays. In a basis, where the right-handed neutrino mass
matrix $M = h_r\VEV{R}$ is diagonal, one obtains \cite{cov,bp2}
\beqa
\ve_1&=&{\Gamma(N_1\rightarrow l \, H_2)-\Gamma(N_1\rightarrow l^c \, H_2^c)
        \over
        \Gamma(N_1\rightarrow l \, H_2)+\Gamma(N_1\rightarrow l^c \, H_2^c)}
        \NO\\[1ex]
     &\simeq&-{3\over16\pi}\;{1\over\left(h_\n h_\n^\dg\right)_{11}}
      \sum_{i=2,3}\mbox{Im}\left[\left(h_\n h_\n^\dg\right)_{1i}^2\right]
      {M_1\over M_i}\label{cpa}\;.
\eeqa
Here we have assumed $M_1 < M_2,M_3$, which is satisfied in the
applications considered in the following sections.  

In the early universe at temperatures $T \sim M_1$ the CP asymmetry 
(\ref{cpa}) leads to a lepton asymmetry\cite{kw},
\beq\label{basym}
Y_L\ =\ {n_L-n_{\Bar{L}}\over s}\ =\ \k\ {\ve_1\over g_*}\;.
\eeq
Here the factor $\k<1$ represents the effect of washout processes. In order
to determine $\k$ one has to solve the full Boltzmann equations 
\cite{lut92,plu97}. In the examples discussed below one has 
$\k\simeq 10^{-1}\ldots 10^{-3}$.

\section{Neutrino masses and mixings}

The CP asymmetry (\ref{cpa}) is given in terms of the Dirac and the Majorana
neutrino mass matrices. Depending on the neutrino mass hierarchy and the
size of the mixing angles the CP asymmetry can vary over many orders of
magnitude. It is therefore interesting to see whether a pattern of neutrino
masses motivated by other considerations is consistent with leptogenesis. 

An attractive framework to explain the observed mass hierarchies of quarks
and charged leptons is the Froggatt-Nielsen mechanism \cite{fro79} based
on a spontaneously broken U(1)$_F$ generation symmetry. The Yukawa couplings 
arise from non-renormalizable interactions after a gauge singlet field $\F$ 
acquires a vacuum expectation value,
\beq
h_{ij} = g_{ij} \left({\VEV\F\over \L}\right)^{Q_i + Q_j}\;.
\eeq
Here $g_{ij}$ are couplings $\co(1)$ and $Q_i$ are the U(1) charges of the
various fermions with $Q_{\F}=-1$. The interaction scale $\L$ is
expected to be very large, $\L > \L_{GUT}$. In the following we shall
discuss two different realizations of this idea which are motivated by
the recently reported atmospheric neutrino anomaly \cite{atm98}. Both
scenarios have a large $\n_\m -\n_\t$ mixing angle. They differ, however,
by the symmetry structure and by the size of the parameter $\e$ which
characterizes the flavour mixing.

\subsection{$SU(5)\times U(1)_F$}
 
This symmetry has been considered by a number of authors \cite{lol99}.
Particularly interesting is the case with a nonparallel family structure
where the chiral $U(1)_F$ charges are different for the $\bf 5^*$-plets
and the $\bf 10$-plets of the same family \cite{sat98}-\cite{bij87}. An
example of possible charges $Q_i$ is given in table~1.

\begin{table}[b]
\begin{center}
\begin{tabular}{c|ccccccccc}\hline \hline
$\j_i$       & $ e^c_{R3}$ & $ e^c_{R2}$  & $ e^c_{R1}$  & $ l_{L3}$    & 
$ l_{L2}$    & $ l_{L1}$   & $ \n^c_{R3}$ & $ \n^c_{R2}$ & $ \n^c_{R1}$ 
\\\hline
$Q_i$  & 0 & 1 & 2 & $0$ & $0$ & $1$ & 0 & $1$ & $2$ \\ \hline\hline
\end{tabular}
\end{center}
\caption{{\it Chiral charges of charged and neutral leptons with
   $SU(5)\times U(1)_F$ symmetry} \cite{buc99}.}
\end{table}

The assignment of the same charge to the lepton doublets of the second and 
third generation leads to a neutrino mass matrix of the form
\cite{sat98,ram98}, 
\beq\label{matrix}
m_{\n_{ij}} \sim \left(\begin{array}{ccc}
    \e^2  & \e  & \e \\[-1ex]
    \e  & \; 1 \; & 1 \\[-1ex]
    \e  &  1  & 1 
    \end{array}\right) {v_2^2\over \VEV R}\;.
\eeq
This structure immediately yields a large $\n_\m -\n_\t$ mixing angle. The
phenomenology of neutrino oscillations depends on the unspecified coefficients
$\co(1)$. The parameter $\e$ which gives the flavour mixing is chosen to be
\beq\label{exp1}
{\VEV\F\over\L} = \e  \sim {1\over 17}\;. 
\eeq
The three Yukawa matrices for the leptons have the structure,
\beq\label{yuk1}
h_e  \sim\ \left(\begin{array}{ccc}
    \e^3 & \e^2 & \e^2 \\[-1ex]
    \e^2 &\;  \e \;   & \e   \\[-1ex]
    \e   & 1    & 1
    \end{array}\right) \;, \quad
h_{\n}  \sim\ \left(\begin{array}{ccc}
    \e^3 & \e^2 & \e^2 \\[-1ex]
    \e^2 &\;  \e \;   & \e   \\[-1ex]
    \e   & 1    & 1
    \end{array}\right) \;, \quad
h_{r}  \sim\ \left(\begin{array}{ccc}
    \e^4 & \e^3 & \e^2 \\[-1ex]
    \e^3 &\;  \e^2 \; & \e   \\[-1ex]
    \e^2 & \e   & 1
    \end{array}\right) \;.
\eeq
Note, that $h_e$ and $h_\n$ have the same, non-symmetric structure.
One easily verifies that the mass ratios for charged leptons, heavy and
light Majorana neutrinos are given by 
\beqa
\qquad\quad 
m_e : m_\m : m_\t \sim \e^3 : \e : 1\;, &\quad&
M_1 : M_2  : M_3  \sim \e^4 : \e^2 : 1\;,\\
m_1 : m_2  : m_3  &\sim& \e^2 : 1 : 1\;.
\eeqa
The masses of the two eigenstates $\n_\m$ and $\n_\t$ depend on unspecified 
factors of order one, and may easily differ by an order of magnitude 
\cite{irg98}. They can therefore be consistent with the mass differences 
$\D m^2_{\n_e \n_\m}\simeq 4\cdot 10^{-6} - 1\cdot 10^{-5}$~eV$^2$ \cite{sol98}
inferred from the MSW solution of the solar neutrino problem \cite{msw86} and 
$\D m^2_{\n_\m \n_\t}\simeq (5\cdot 10^{-4}-6\cdot 10^{-3})$~eV$^2$ associated 
with the atmospheric neutrino deficit \cite{atm98}. For numerical estimates 
we shall use the average of the neutrino masses of the second and third family,
$\Bar{m}_\n=(m_{\n_\m}m_{\n_\t})^{1/2} \sim 10^{-2}$~eV. Note, that for a
different choice of U(1) charges the coefficients in eq.~(\ref{matrix}) 
automatically yield the hierarchy $m_2/m_3 \sim \e^{2/3}$ \cite{alt98}.

\begin{table}[b]
\begin{center}
\begin{tabular}{c|ccccccccc}
\hline \hline
$\j_i$       & $ e^c_{R3}$ & $ e^c_{R2}$  & $ e^c_{R1}$  & $ l_{L3}$    & 
$ l_{L2}$    & $ l_{L1}$   & $ \n^c_{R3}$ & $ \n^c_{R2}$ & $ \n^c_{R1}$ 
\\\hline
$Q_i$  & 0 & ${1\over 2}$ & ${5\over 2}$ & $0$ & ${1\over 2}$ & ${5\over 2}$ 
& 0 & ${1\over 2}$ & ${5\over 2}$ \\ \hline\hline
\end{tabular}
\end{center}
\caption{{\it Chiral charges of charged and neutral leptons with
$SU(3)_c \times SU(3)_L \times SU(3)_R \times U(1)_F$ symmetry}
\cite{lol99}.}
\end{table}

The choice of the charges in table~1 corresponds to large Yukawa couplings
of the third generation. For the mass of the heaviest Majorana neutrino
one finds
\beq
M_3\ \sim\ {v_2^2\over\Bar{m}_\n}\ \sim\ 10^{15}\ \mbox{GeV}\;.
\eeq
This implies that $B-L$ is broken at the unification scale $\L_{GUT}$.

\subsection{$SU(3)_c \times SU(3)_L \times SU(3)_R \times U(1)_F$}

This symmetry arises in unified theories based on the gauge group $E_6$.
The leptons $e_R^c$, $l_L$ and $\n_R^c$ are contained in a single
$(1,3,\bar{3})$ representation. Hence, all leptons of the same generation
have the same $U(1)_F$ charge and all leptonic Yukawa matrices are
symmetric. Masses and mixings of quarks and charged leptons can be
successfully described by using the charges given in table~2 \cite{lol99}. 
Clearly, the three Yukawa matrices have the same structure\footnote{Note,
that with respect to ref.~\cite{lol99}, $\e$ and $\Bar{\e}$ have been
interchanged.},
\beq\label{yuk2}
h_e,\ h_r  \sim\ \left(\begin{array}{ccc}
    \e^5        & \e^3        & \e^{5/2} \\
    \e^3        & \e          & \e^{1/2} \\
    \e^{5/2} \; & \e^{1/2} \; & 1
    \end{array}\right) \;, \quad
h_{\n}  \sim\ \left(\begin{array}{ccc}
    \bar\e^5        &\bar\e^3        & \bar\e^{5/2} \\
    \bar\e^3        &\bar\e          & \bar\e^{1/2} \\
    \bar\e^{5/2} \; &\bar\e^{1/2} \; & 1
    \end{array}\right) \;.
\eeq
Note, that the expansion parameter in $h_{\n}$ is different from the one in
$h_e$ and $h_r$. From the quark masses, which also contain $\e$ and
$\bar{\e}$, one infers $\bar{\e} \simeq \e^2$ \cite{lol99}.

From eq.~(\ref{yuk2}) one obtains for the masses of charged leptons,
light and heavy Majorana neutrinos,
\beq
m_e : m_\m : m_\t\ \sim\ M_1 : M_2  : M_3\ \sim \e^5 : \e : 1\;, 
\eeq
\beq
m_1 : m_2 : m_3 \ \sim\ \e^{15} : \e^3 : 1\;.
\eeq
Like in the example with $SU(5)\times U(1)_F$ symmetry, the mass of the
heaviest Majorana neutrino,
\beq
M_3 \sim {v_2^2\over m_3} \sim 10^{15}\;\mbox{GeV} \;,
\eeq
implies that $B-L$ is broken at the unification scale $\L_{GUT}$.

The $\n_\m-\n_\t$ mixing angle is mostly given by the mixing of
the charged leptons of the second and third generation \cite{lol99},
\beq
\sin{\Q_{\m\t}} \sim \sqrt{\e} + \e \;.
\eeq
This requires large flavour mixing,
\beq\label{exp2}
\left({\VEV\F\over\L}\right)^{1/2} = \sqrt{\e}  \sim {1\over 2}\;. 
\eeq 
In view of the unknown coefficients $\co(1)$ the corresponding mixing angle 
$\sin{\Q_{\m\t}} \sim 0.7$ is consistent with the interpretation of the 
atmospheric neutrino anomaly as $\n_\m-\n_\t$ oscillation.

It is very instructive to compare the two scenarios of lepton masses and
mixings described above. In the first case, the large $\n_\m-\n_\t$
mixing angle follows from a nonparallel flavour symmetry. The parameter $\e$,
which characterizes the flavour mixing, is small. In the second case, the
large $\n_\m-\n_\t$ mixing angle is a consequence of the large flavour
mixing $\e$. The $U(1)_F$ charges of all leptons are the same, i.e., one
has a parallel family structure. Also the mass hierarchies, given in terms
of $\e$, are rather different. This illustrates that the separation into
a flavour mixing parameter $\e$ and coefficients $\co(1)$ is far from
unique. It is therefore important to study other observables which
depend on the lepton mass matrices. A particular example is the baryon
asymmetry.

\section{Matter antimatter asymmetry}

We can now evaluate the baryon asymmetry for the two patterns of neutrino
mass matrices discussed in the previous section. A rough estimate of the
baryon asymmetry can be obtained from the CP asymmetry $\ve_1$ of the heavy
Majorana neutrino $N_1$. A quantitative determination requires a numerical
study of the full Boltzmann equations \cite{plu97}.

\subsection{$SU(5)\times U(1)_F$}

In this case one obtains from eqs.~(\ref{cpa}) and (\ref{yuk1}),
\beq
\ve_1\ \sim\ {3\over 16\pi}\ \e^4\;.
\eeq 
From eq.~(\ref{basym}), $\e^2 \sim 1/300$ (\ref{exp1}) and $g_* \sim 100$ 
one then obtains the baryon asymmetry,
\beq\label{est1}
Y_B \sim \k\ 10^{-8}\;.
\eeq
For $\k \sim 0.1\ldots 0.01$ this is indeed the correct order of magnitude.
The baryogenesis temperature is given by the mass of the lightest of the
heavy Majorana neutrinos,
\beq
T_B \sim M_1 \sim \e^4 M_3 \sim 10^{10}\ \mbox{GeV}\;.
\eeq
This set of parameters, where the CP asymmetry is given in terms of the mass 
hierarchy of the heavy neutrinos, has been studied in detail \cite{buc96}. 
The generated baryon asymmetry does not depend on the flavour mixing of the 
light neutrinos. The $\n_\m-\n_\t$ mixing angle is large in the scenario 
described in the previous section whereas it was assumed to be small in 
\cite{buc96}.

 \begin{figure}[t]
    \mbox{ }\hfill
    \epsfig{file=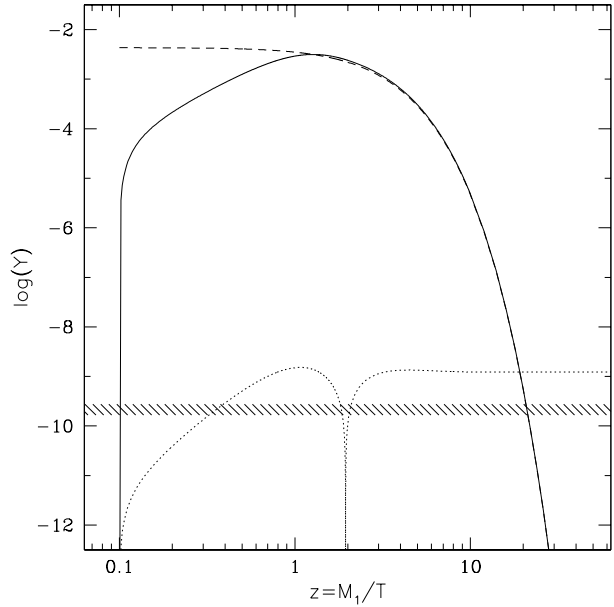,width=8.2cm}
    \hfill\mbox{ }
    \caption{\it Time evolution of the neutrino number density and the
     lepton asymmetry in the case of the $SU(5)\times U(1)_F$ symmetry. 
     The solid line shows the solution of the Boltzmann equation for the 
     right-handed neutrinos, while the corresponding equilibrium 
     distribution is represented by the dashed line.
     The absolute value of the lepton asymmetry $Y_L$ 
     is given by the dotted line and the hatched area shows the
     lepton asymmetry corresponding to the observed baryon asymmetry.
     \label{asyB}}
  \end{figure} 

The solution of the full Boltzmann equations is shown in fig.~\ref{asyB} 
for the non-supersymmetric case \cite{buc96}. The initial condition at a 
temperature $T \sim 10 M_1$ is chosen to be a state without heavy neutrinos. 
The Yukawa interactions are sufficient to bring the heavy neutrinos into 
thermal equilibrium. At temperatures $T\sim M_1$ this is followed by the usual 
out-of-equilibrium decays which lead to a non-vanishing baryon asymmetry. 
The final asymmetry agrees with the estimate (\ref{est1}) for $\k \sim 0.1$. 

The change of sign in the lepton asymmetry is due to the fact that
inverse decay processes, which take part in producing the neutrinos, 
are $CP$ violating, i.e.\ they generate a lepton 
asymmetry at high temperatures. Due to the interplay of inverse decay
processes and lepton number violating scattering processes this asymmetry
has a different sign than the one produced by neutrino decays at lower
temperatures.

\subsection{$SU(3)_c\times SU(3)_L\times SU(3)_R\times U(1)_F$}

\begin{figure}[t]
    \mbox{ }\hfill
    \epsfig{file=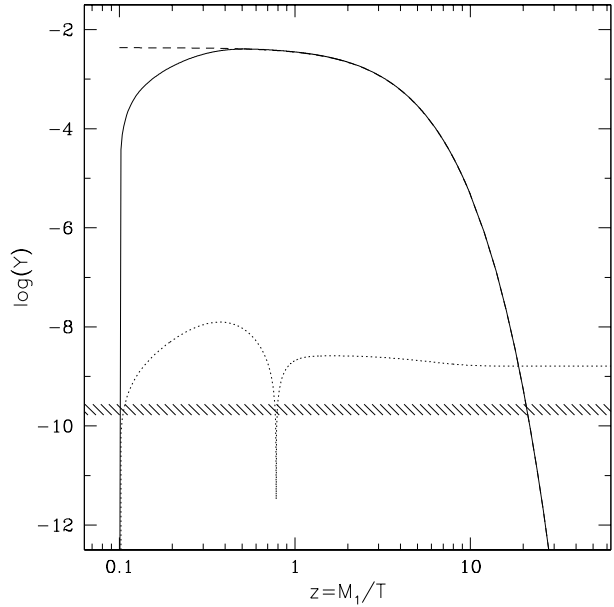,width=8.2cm}
    \hfill\mbox{ }
    \caption{\it Solution of the Boltzmann equations in the case of the
     $SU(3)_c\times SU(3)_L\times SU(3)_R\times$ $U(1)_F$ symmetry. 
     \label{asyLR}}
  \end{figure} 

In this case the neutrino Yukawa couplings (\ref{yuk2}) yield the CP asymmetry
\beq
\ve_1\ \sim\ {3\over 16\pi}\ \e^5\;,
\eeq 
which correspond to the baryon asymmetry (cf.~(\ref{basym}))
\beq\label{est2}
Y_B \sim \k\ 10^{-6}\;.
\eeq
Due to the large value of $\e$ the CP asymmetry is two orders of magnitude
larger than in the case with $SU(5)\times U(1)_F$ symmetry. However, 
washout processes are now also stronger. The solution of the Boltzmann 
equations is shown in fig.~\ref{asyLR}. The final asymmetry is again 
$Y_B \sim 10^{-9}$ 
which now corresponds to $\k \sim 10^{-3}$. The baryogenesis temperature is
considerably larger than in the first case,
\beq
T_B \sim M_1 \sim \e^5 M_3 \sim 10^{12}\ \mbox{GeV}\;.
\eeq

The baryon asymmetry is largely determined by the parameter $\wt m_1$
defined in eq.~(\ref{ooeb}) \cite{plu97}. In the first example, one
has $\wt m_1 \sim \Bar m_\n$. In the second case one finds  
$\wt m_1 \sim m_3$. Since $\Bar m_\n$ and $m_3$ are rather similar
it is not too surprizing that the generated baryon asymmetry is about
the same in both cases.

\section{Conclusions}

Detailed studies of the thermodynamics of the electroweak interactions at
high temperatures have shown that in the standard model and most of its
extensions the electroweak transition is too weak to affect the 
cosmological baryon asymmetry. Hence, one has to search for baryogenesis
mechanisms above the Fermi scale. 

Due to sphaleron processes baryon number and lepton number are related
in the high-temperature, symmetric phase of the standard model. As a
consequence, the cosmological baryon asymmetry is related to neutrino
properties. Baryogenesis requires lepton number violation, which occurs
in extensions of the standard model with right-handed neutrinos and
Majorana neutrino masses. 

Although lepton number violation is needed in order to obtain a baryon
asymmetry, it must not be too strong since otherwise any baryon and lepton
asymmetry would be washed out. This leads to stringent upper bounds on
neutrino masses which depend on the particle content of the theory.

The solar and atmospheric neutrino deficits can be interpreted as a result
of neutrino oscillations. For hierarchical neutrinos the corresponding
neutrino masses are very small. Assuming the see-saw mechanism, this suggests
the existence of very heavy right-handed neutrinos and a large scale of
$B-L$ breaking.

It is remarkable that these hints on the nature of lepton number violation
fit very well together with the idea of leptogenesis.
For hierarchical neutrino masses, with $B-L$ broken at the
unification scale $\Lambda_{\mbox{\scriptsize GUT}}\sim 10^{16}\;$GeV,  
the observed baryon asymmetry $Y_B \sim 10^{-10}$ is naturally
explained by the decay of heavy Majorana neutrinos. 

Although the observed baryon asymmetry imposes important constraints on
neutrino properties, other observables are needed to discriminate
between different models. The two examples considered in this paper
predict different baryogenesis temperatures. Correspondingly, in
supersymmetric models the predictions for the gravitino abundance are
different \cite{khl84}-\cite{bol98}. In the case with $SU(5)\times U(1)_F$
symmetry, stable gravitinos can be the dominant component of cold dark
matter \cite{bol98}. The models make also different predictions for the
rate of lepton flavour changing radiative corrections. 

%\clearpage
%\mbox{ }\vspace{4ex}\\
%\noindent
%\setlength{\parskip}{1ex}
%{\bf Acknowledgments}\\
%\mbox{ }\\    
%The author would like to thank 
%\mbox{ }\\\noindent


\begin{thebibliography}{99}

\bibitem{sac}
A.~D.~Sakharov, JETP Lett.~{\bf 5} (1967) 24
%%CITATION = ZFPRA,5,32;%%

\bibitem{thoo}
G.~'t~Hooft, \prl{37}{76}{8}
%%CITATION = PRLTA,37,8;%%

\bibitem{krs}
V.~A.~Kuzmin, V.~A.~Rubakov, M.~E.~Shaposhnikov, \pl{155}{85}{36}
%%CITATION = PHLTA,155B,36;%%

\bibitem{chem}
J.~A.~Harvey, M.~S.~Turner, \pr{42}{90}{3344}
%%CITATION = PHRVA,D42,3344;%%

\bibitem{fy1}
M.~Fukugita, T.~Yanagida, \pr{42}{90}{1285}
%%CITATION = PHRVA,D42,1285;%%

\bibitem{cli93}
J.~M.~Cline, K.~Kainulainen, K.~A.~Olive, \prl{71}{93}{2372}
%%CITATION = %%

\bibitem{dol}
For a review and references, see\\
A.~D.~Dolgov, \prep{222}{92}{309};\\
%%CITATION = PRPLC,222,309;%%
V.~A.~Rubakov, M.~E.~Shaposhnikov, Phys.~Usp.~{\bf 39} (1996) 461;\\
%%CITATION = UFNAA,166,493;%%
S.~J.~Huber, M.~G.~Schmidt, {\it SUSY Variants of the Electroweak Phase
Transition}, {\tt hep-ph/9809506}
%%CITATION = HEP-PH 9809506;%%

\bibitem{jansen} 
For a discussion and references, see \\ 
K.~Jansen, Nucl.~Phys.~B (Proc.~Supp.) 47 (1996) 196;\\
%%CITATION = NUPHZ,47,196;%%
W.~Buchm\"uller, in {\it Quarks '96} (Yaroslavl, Russia, 1996) eds.
V.~A.~Matveev et al., {\tt hep-ph/9610335};\\
%%CITATION = HEP-PH 9610335;%%
K.~Rummukainen, Nucl.~Phys.~B (Proc.~Suppl.) {\bf 53} (1997) 30
%%CITATION = NUPHZ,53,30;%%

\bibitem{seesaw} 
T.~Yanagida, in {\it{Workshop on unified Theories}}, KEK report 
79-18 (1979) p.~95;\\
M.~Gell-Mann, P.~Ramond, R.~Slansky, in {\it{Supergravity}} (North Holland, 
Amsterdam, 1979) eds. P.~van Nieuwenhuizen, D.~Freedman, p.~315

\bibitem{fy} 
M.~Fukugita, T.~Yanagida, \pl{174}{86}{45}
%%CITATION = PHLTA,174B,45;%%

\bibitem{fisch}
W.~Fischler, G.~F.~Giudice, R.~G.~Leigh, S.~Paban, \pl{258}{91}{45}
%%CITATION = PHLTA,B258,45;%%

\bibitem{cov}
L.~Covi, E.~Roulet, F.~Vissani, \pl{384}{96}{169};\\
%%CITATION = PHLTA,B384,169;%%
M.~Flanz, E.~A.~Paschos, U.~Sarkar, \pl{345}{95}{248}; \pl{384}{96}{487} (E)
%%CITATION = PHLTA,B345,248;%%

\bibitem{bp2}
W.~Buchm\"uller, M.~Pl\"umacher, \pl{431}{98}{354}
%%CITATION = PHLTA,B431,354;%%

\bibitem{kw} 
A.~D.~Dolgov, Ya.~B.~Zeldovich, Rev.~Mod.~Phys.~{\bf 53} (1981) 1;\\
%%CITATION = RMPHA,53,1;%%
E.~W.~Kolb, S.~Wolfram, \np{172}{80}{224}; \np{195}{82}{542}(E)
%%CITATION = NUPHA,B172,224;%%

\bibitem{lut92}
M.~A.~Luty, \pr{45}{92}{455}
%%CITATION = PHRVA,D45,455;%%

\bibitem{plu97}
M.~Pl\"umacher, Z.~Phys.~{\bf C\ 74} (1997) 549;
%%CITATION = ZEPYA,C74,549;%%
\np{530}{98}{207}
%%CITATION = NUPHA,B530,207;%%

\bibitem{atm98}
Super-Kamiokande Collaboration, Y.~Fukuda et al., \prl{81}{98}{1562}
%%CITATION = PRLTA,81,1562;%%

\bibitem{fro79}
C.~D.~Froggatt, H.~B.~Nielsen, \np{147}{79}{277}
%%CITATION = NUPHA,B147,277;%%

\bibitem{sat98}
J.~Sato, T.~Yanagida, Talk at {\it Neutrino'98}, {\tt hep-ph/9809307}
%%CITATION = HEP-PH 9809307;%%

\bibitem{ram98}
P.~Ramond, Talk at {\it Neutrino'98}, {\tt hep-ph/9809401}
%%CITATION = HEP-PH 9809401;%%

\bibitem{bij87}
J.~Bijnens, C.~Wetterich, \np{292}{87}{443}
%%CITATION = NUPHA,B292,443;%%

\bibitem{lol99}
For a recent discussion and references, see\\
S.~Lola, G.~G.~Ross, {\tt hep-ph/9902283} 
%%CITATION = HEP-PH 9902283;%%

\bibitem{buc99}
W.~Buchm\"uller, T.~Yanagida, \pl{445}{99}{399}
%%CITATION = PHLTA,B445,399;%%

\bibitem{vis98}
F.~Vissani, JHEP11 (1998) 025
%%CITATION = JHEPA,9811,025;%%

\bibitem{irg98}
N.~Irges, S.~Lavignac, P.~Ramond, \pr{58}{98}{035003}
%%CITATION = PHRVA,D58,035003;%%

\bibitem{buc96}
W.~Buchm\"uller, M.~Pl\"umacher, \pl{389}{96}{73}
%%CITATION = PHLTA,B389,73;%%

\bibitem{sol98}
N.~Hata, P.~Langacker, \pr{56}{97}{6107}
%%CITATION = PHRVA,D56,6107;%%

\bibitem{msw86}
S.~P.~Mikheyev, A.~Y.~Smirnov, \nc{9C}{86}{17};\\
%%CITATION = NUCIA,9C,17;%%
L.~Wolfenstein, \pr{17}{78}{2369}
%%CITATION = PHRVA,D17,2369;%%

\bibitem{alt98}
G.~Altarelli, F.~Feruglio, JHEP11 (1998) 021; {\tt hep-ph/9812475}
%%CITATION = JHEPA,9811,021;%%

\bibitem{khl84}
M.~Yu.~Khlopov, A.~D.~Linde, \pl{138}{84}{265};\\
%%CITATION = PHLTA,138B,265;%%
J.~Ellis, J.~E.~Kim, D.~V.~Nanopoulos, \pl{145}{84}{181}
%%CITATION = PHLTA,145B,181;%%

\bibitem{mor}
T.~Moroi, H.~Murayama, M.~Yamaguchi, \pl{303}{93}{289}
%%CITATION = PHLTA,B303,289;%%

\bibitem{bol98}
M.~Bolz, W.~Buchm\"uller, M.~Pl\"umacher, \pl{443}{98}{209}
%%CITATION = PHLTA,B443,209;%%

\end{thebibliography}
\end{document}